\documentclass[osajnl,twocolumn,showpacs,superscriptaddress,10pt]{revtex4-1} 
\usepackage{amsmath,amssymb,graphicx}
\begin{document}

\title{Direct Experimental  Simulation of the Yang-Baxter Equation }

\author{Chao Zheng}
\affiliation{State Key Laboratory of Low-Dimensional Quantum Physics and Department of Physics, Tsinghua University, Beijing 100084, P.R. China}
\affiliation{Tsinghua National Laboratory for Information Science and Technology, Beijing 100084, P. R. China},

\author{Jun-lin Li}
\affiliation{State Key Laboratory of Low-Dimensional Quantum Physics and Department of Physics, Tsinghua University, Beijing 100084, P.R. China},

\author{Si-yu Song}
\affiliation{State Key Laboratory of Low-Dimensional Quantum Physics and Department of Physics, Tsinghua University, Beijing 100084, P.R. China}

\author{Gui Lu Long}\email{Corresponding author: gllong@tsinghua.edu.cn}
\affiliation{State Key Laboratory of Low-Dimensional Quantum Physics and Department of Physics, Tsinghua University, Beijing 100084, P.R. China}
\affiliation{Tsinghua National Laboratory for Information Science and Technology, Beijing 100084, P. R. China}

\begin{abstract}
     Introduced in the field of many-body statistical mechanics,  Yang-Baxter equation has become an important tool in a variety fields of physics. In this work, we report the first direct experimental simulation of the Yang-Baxter equation using linear quantum optics.  The equality between the two sides of the Yang-Baxter equation in two dimension has been demonstrated directly, and the spectral parameter transformation in the Yang-Baxter equation is explicitly confirmed.

\end{abstract}


\maketitle

\section{Introduction}

Yang-Baxter equation (YBE) was originated from solving the
repulsive $\delta$ interaction problem in one-dimension of $N$
particles \cite{th1,th1-1},
and problems of statistical models on lattices \cite{th2,th2-2,th3}.
Today, Yang-Baxter equation has become an important tool in physics, and has many applications in variety of areas of physics,
for instance in quantum field theory, statistical mechanics, and group theories
\cite{th2,th2-2,th3,thpnas1,thpnas2,th4,thpnas3,ybe11,ybe12,ybe13,ybe14,ybe15,ap1}. It can be
applied in completely integrable statistical models to find the
solutions by means of the nested Bethe ansatz \cite{ap1}. Recently
it turns up gradually that Yang-Baxter equation is naturally linked
to a hot area of frontier research, the quantum information and
computing \cite{nielsen,long}.
It is found that  the Yang-Baxter equation is closely
related to quantum entangled states \cite{ap6,ap7}, the braiding operations in the Yang-Baxter equation are universal quantum
gates \cite{ap2-2,ap3,ap4,ap5,ap8}. Yang-Baxter equation attracts much attention in recent years and is being studied  in the context of quantum correlation and entanglement, and
topological quantum computing intensively\cite{ap2,ap2-1,ap9,ap9-1,ap9-2,ap9-3,ap10,panjw}.

Due to its importance, the experimental verification of Yang-Baxter
equation has been pursued all along. Notably, an experimental verification
was carried out by Tennant et al in 1995 \cite{intest0,intest1}. Tennant
et al measured the spectrum of Heisenberg spin-half chain, and the
experimental result appeared to agree with the calculation based on
the Yang-Baxter equation. Recently, the density profile of 1-dimensional wires was measured
and it agreed well with the theoretical calculations
based on Yang's solvable model \cite{intest2}. However, these experiments
are indirect verifications of the Yang-Baxter equation because
the Yang-Baxter equation provides only a sufficient condition for
the spectrum or profile, or the observed profile is only a necessary condition for the Yang-Baxter equation and it does not guarantee the validity of the Yang-Baxter equation.  Thus the direct verification of the Yang-Baxter equation is still an open question \cite{opttest}.
Direct experimental verification of the Yang-Baxter equation requires
not only the  verification of the equality of the left-hand and the right-hand
sides of the equation, but also the transformation relation between
the spectral parameters in the Yang-Baxter equation, the Lorentz-like
transformation.

In this paper, we report the first direct experimental simulation
of the Yang-Baxter equation using quantum optics. The fundamental principles of the present simulation was established in 2008 by Hu, Xue and Ge \cite{opttest}.  Hu, Xue and Ge gave an
explicit optical realization of the Yang-Baxter equation.  By the use of the Temperley-Lieb algebra, they made a remarkable reduction that obtained a Yang-Baxter equation with  dimension 2, the minimum dimensional Yang-Baxter equation so far.  This makes it possible to be implemented in quantum
optics with current technology. In our experiment, we experimentally implemented the Hu-Xue-Ge scheme and demonstrated the validity of the Yang-Baxter equation using linear quantum optical
components such as beamsplitters, half-wave plates, quarter wave
plates, and etc. The equality of the two sides of the Yang-Baxter
equation is directly verified. In particular, the Lorentz-like
transformation in the spectral parameters of the Yang-Baxter
equation is experimentally demonstrated for the first time. The
present experiment completes the first direct experimental
simulation of the Yang-Baxter equation.


\section{Theoretical framework}

The Yang-Baxter equation reads, 
\begin{eqnarray}
\breve{R}_{12}(u)\breve{R}_{23}(u_{23})\breve{R}_{12}(v)=\breve{R}_{23}(v)\breve{R}_{12}(u_{23})\breve{R}_{23}(u),\label{e0}
\end{eqnarray}
 where $u$ and $v$ are spectral parameters, and $\beta^{-1}=ic$
($c$ is the light speed in vacuum), and
\begin{eqnarray}
u_{23}=\frac{u+v}{1+\beta uv},
\end{eqnarray}
 is the Lorentz-like transformation relation of the spectral parameters.
The $N^{2}\times N^{2}$ dimension matrix $\breve{R}$ acts on the
tensor product space $V\otimes V$ of two $N$-dimensional spaces, and is the two-particle scattering matrix
depending on the relative rapidity $\tanh^{-1}(\beta u)$. When $\beta u=1$, $\breve{R}=b$ which is a braid matrix, and the Yang-Baxter equation reduces
to the braid relation $b_{12}b_{23}b_{12}= b_{23} b_{12} b_{23}$.  This equation implies the scattering of particles 1 and 2, followed by scattering of particles 2 and 3, and then scattering of particles 1 and 2, is equal to the scattering of particles 2 and 3, followed by scattering of particles 1 and 2, and then scattering of particles 2 and 3, when they satisfy the Yang-Baxter equation with suitable spectral parameters.

Yang-Baxter equation is an abstract equation and the quantities in the equation may have different meanings in different problem. For instance, it has been found recently that the braid matrix and the Yang-Baxter equation are connected to entangled quantum states \cite{ap3}. The Bell-basis entangled states in four-dimension can be obtained by applying braid operation that satisfies the Yang-Baxter equation on the computational basis. Here the matrix in the Yang-Baxter equation becomes a transformation that transforms the computational basis to the Bell-basis states. There have been active studies in this direction,   interested readers can refer to Ref. \cite{opttest} and references therein for more details.

The scheme used in our experiment was proposed by Hu, Xue and Ge
recently \cite{opttest}. The Yang-Baxter equation with the minimum
nontrivial dimension is in four dimension. In this case, $\breve{R}$
becomes a $4\times4$ matrix. In principle, it can be simulated
directly by means of quantum optics, and Hu, Xue and Ge gave an
explicit optical realization. However, such a realization requires
many controlled NOT gates whose realization is of very low
efficiency in linear quantum optics \cite{cnot1,cnot2} that its
feasibility using current technology is illusive. A further
simplification was made by the use of the Temperley-Lieb algebra
\cite{tla}, and the Yang-Baxter equation with minimal dimension was
reduced further to 2. This makes it feasible to implement in quantum
optics with current technology.

The 2-dimensional Yang-Baxter equation is expressed as,
\begin{equation}
A(u)B(\frac{u+v}{1+\beta^{2}uv})A(v)=B(v)A(\frac{u+v}{1+\beta^{2}uv})B(u),\label{e1}
\end{equation}
 where
\begin{equation}
A(u)=\rho(u)\left(\begin{array}{cc}
\frac{1+\beta^{2}u^{2}+2i\epsilon\beta u}{1+\beta^{2}u^{2}-2i\epsilon\beta u} & 0\\
0 & 1
\end{array}\right),\label{e2}
\end{equation}
 and
\begin{equation}
B(u)=\frac{\rho(u)}{1+\beta^{2}u^{2}-2i\epsilon\beta
u}\left(\begin{array}{cc}
1+\beta^{2}u^{2} & 2i\epsilon\beta u\\
2i\epsilon\beta u & 1+\beta^{2}u^{2}
\end{array}\right),\label{e3}
\end{equation}
 and $\rho(u)$ is a normalization factor and $\epsilon=\pm1$.

For convenience in optical realization, $A$ and $B$ are represented
as functions of optical parameter $\theta$, the angle between the
optical axes of an optical device and the vertical direction. The
two sets of parameters are related by using the following
transformation
\begin{equation}
\frac{1+\beta^{2}u^{2}+2i\epsilon\beta
u}{1+\beta^{2}u^{2}-2i\epsilon\beta u}\equiv e^{-2i\theta}\label{e4}
\end{equation}
 and
\begin{equation}
\rho(u)\equiv e^{i\theta}.\label{e5}
\end{equation}
 $A(u)$ and $B(u)$ then become simple matrices in two dimensions
\begin{equation}
A(\theta)=\left(\begin{array}{cc}
e^{-i\theta} & 0\\
0 & e^{i\theta}
\end{array}\right),\label{e6}
\end{equation}
 and
\begin{equation}
B(\theta)=\left(\begin{array}{cc}
\cos\theta & -i\,\sin\theta\\
-i\,\sin\theta & \cos\theta
\end{array}\right).\label{e7}
\end{equation}
 The Yang-Baxter equation in Eq. (\ref{e1}) can be re-written as
\begin{equation}
A(\theta_{1})B(\theta_{2})A(\theta_{3})=B(\theta_{3})A(\theta_{2})B(\theta_{1}).\label{e8}
\end{equation}
 The three parameters in the Yang-Baxter equation $\theta_{1}$, $\theta_{2}$
and $\theta_{3}$ are not independent, and they are related through
the following equation
\begin{equation}
(e^{-2i\theta_{2}}+1)[i-\sec(\theta_{1}-\theta_{3})\sin(\theta_{1}+\theta_{3})]=2i.\label{e9}
\end{equation}
 Using this relation, we can transform the Lorentz-like relation in
spectral parameters into a relation in the optical angle parameters,
\begin{equation}
\theta_{2}=\arctan\left(\frac{\sin(\theta_{1}+\theta_{3})}{\cos(\theta_{1}-\theta_{3})}\right).\label{e10}
\end{equation}

We use the photon polarization qubit in our experiment. A general
elliptically polarized photon state $|\psi\rangle$ can be written as
\begin{equation}
|\psi\rangle=\alpha|\updownarrow\rangle+i\beta|\leftrightarrow\rangle,\label{e11}
\end{equation}
 where $\alpha$ and $\beta$ are real and satisfy $|\alpha|^{2}+|\beta|^{2}=1$.
Without loss of generality, we assume $\alpha$ is real and positive.
The sign of the $\beta$ specifies the handedness of the circular
polarized photon. $|\updownarrow\rangle$ and
$|\leftrightarrow\rangle$ are basis states of the vertical and
horizontal polarization respectively. State $|\psi\rangle$ is
measured directly in the experiment.

\begin{figure}[h]
 \centerline{\includegraphics[scale=0.5]{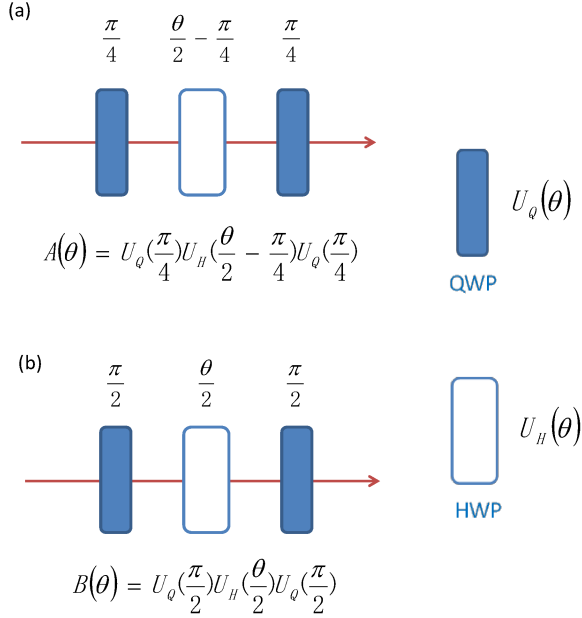}} \caption{(Color online) Realization of operations (a) $A(\theta)$ and $B(\theta)$ by optical
elements.$U_{Q}(\theta)$ and $U_{H}(\theta)$ are the matrices of QWP
and HWP, respectively, and $\theta$ is the angle between the optical
device axes and the vertical direction.}\label{f1}
\end{figure}

The operations of $A(\theta)$ and $B(\theta)$ can be realized by
series of quarter-wave plates (QWP) and half-wave plates (HWP),
whose effects are equivalent to two elements of an SU(2)
transformation group \cite{wp} as shown in Fig. \ref{f1},

\begin{figure}[h]
\centerline{ \includegraphics[scale=0.51]{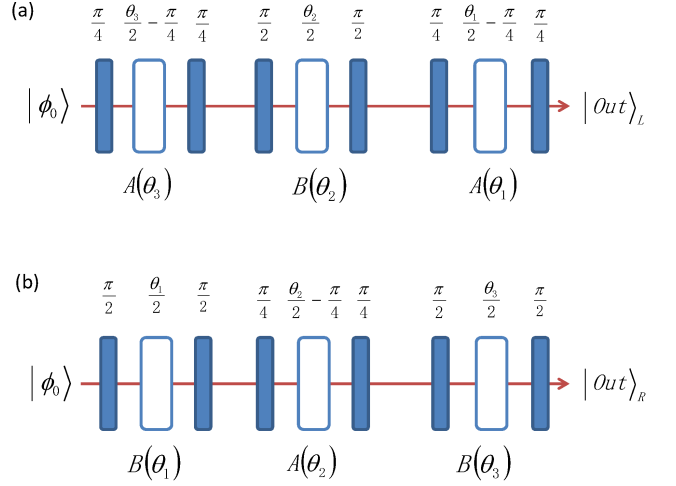}} \caption{(Color online) Optical
realization of the (a) LHS and RHS of Yang-Baxter equation. The
angles of these QWPs (filled) and HWPs (empty) must satisfy the
Lorentz-like relation given in Eq. (\ref{e10}).}\label{f2}
\end{figure}

\begin{equation}
U_{Q}(\theta)=\frac{1}{\sqrt{2}}\left(\begin{array}{cc}
1-i\,\cos(2\theta) & -i\,\sin(2\theta)\\
-i\,\sin(2\theta) & 1+i\,\cos(2\theta)
\end{array}\right)\label{e14}
\end{equation}
 and
\begin{equation}
U_{H}(\theta)=U_{Q}^{2}(\theta)=-i\left(\begin{array}{cc}
\cos(2\theta) & \sin(2\theta)\\
\sin(2\theta) & -\cos(2\theta)
\end{array}\right),\label{e15}
\end{equation}
 where $\theta$ is the angle between the optical axis of QWP or HWP
and the vertical direction. Thus, $A(\theta)$ and $B(\theta)$ can be
simulated by series of QWP and HWP, i.e.
\begin{equation}
A(\theta)=U_{Q}\left(\frac{\pi}{4}\right)U_{H}\left(-\frac{\pi}{4}+\frac{\theta}{2}\right)U_{Q}\left(\frac{\pi}{4}\right)\label{e16}
\end{equation}
 and
\begin{equation}
B(\theta)=U_{Q}\left(\frac{\pi}{2}\right)U_{H}\left(\frac{\theta}{2}\right)U_{Q}\left(\frac{\pi}{2}\right),\label{e17}
\end{equation}
 respectively. The two sides of the Yang-Baxter equation are simulated
by two series of wave plates as illustrated in Fig. \ref{f2}.


The qubit state after the transformation of the LHS of the
Yang-Baxter equation is denoted as,
\begin{equation}
|\psi_{{\rm
out}}\rangle_{L}=\alpha_{L}|\updownarrow\rangle_{L}+i\beta_{L}|\leftrightarrow\rangle_{L}\label{e18}
\end{equation}
 and the qubit state after the processing of the RHS of the Yang-Baxter
equation can be expressed as
\begin{equation}
|\psi_{{\rm
out}}\rangle_{R}=\alpha_{R}|\updownarrow\rangle_{R}+i\beta_{R}|\leftrightarrow\rangle_{R}.\label{e19}
\end{equation}
 To check the equality of these two final output states, we define
the fidelity
\begin{equation}
{\rm C}_{{\rm YBE}}=|{}_{L}\langle\psi_{{\rm out}}|\psi_{{\rm
out}}\rangle_{R}|,\label{e21}
\end{equation}
 which is the absolute value of the overlap of the two states. Fidelity
${\rm C}_{{\rm YBE}}$ is a good measure of the validity of the
Yang-Baxter equation. If ${\rm C}_{{\rm YBE}}$ equals to 1, the two
sides of the Yang-Baxter equation is equal. Otherwise it is not
valid. In the real experiment, it should be 1 within statistical
errors, and independent of the input state.

\section{Experimental method and results}
\begin{figure}[h]
\centerline{\includegraphics[scale=0.3]{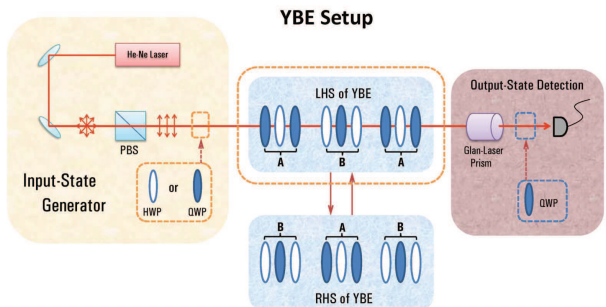}}
\caption{(Color online) Experimental setup. The left part generates a horizontally
polarized state, which is then transformed into a desired input
state with arbitrary linearly or elliptically polarization by using
a HWP or QWP, respectively, following the PBS. The middle part is
the left-hand side or right-hand side of Yang-Baxter equation
consists of series of wave plates. The right part, containing a
Glan-Laser prism and a single photon detector, is used to detect the
polarized state of the output state. A QWP may be inserted to
determine the handedness of an elliptically polarized photon. }
\label{f3}
\end{figure}

The experimental setup is explicated in Fig. \ref{f3}. A He-Ne laser
with center frequency 632.8nm, drawn in the left part of Fig.
\ref{f3}, is used to generate sequences of photons with certain
polarization state. The input state can be prepared at arbitrary
linearly or elliptically polarized photon states conveniently by
using either a HWP or a QWP following the PBS and an attenuator. In
the experiments, the intensity of the light source are attenuated to
a weak level that well approximate the single photon sources. Then
the photons go through a series of optical components which simulate
either the left-hand-side or the right-hand-side of the Yang-Baxter
equation, which is indicated in the middle part of Fig. \ref{f3}.
The right part of Fig. \ref{f3} is used to measure the polarized
state of the photon output state after going through corresponding
Yang-Baxter equation transformation.

\subsection{Verification of Equality of the Yang-Baxter equation}
Groups of angles $\theta_{1}$, $\theta_{2}$ and $\theta_{3}$ are
selected and experimented. The fidelity ${\rm C}_{{\rm YBE}}$
parameter is measured for each group of angles. The experimental
results prove the Yang-Baxter equation very well. For illustration
purpose, we set the two angles $\theta_{1}$ and $\theta_{3}$ at
values $56\textrm{\ensuremath{^{\textrm{o}}}}$ and
$23{}^{\textrm{o}}$ respectively in the following. We then vary
$\theta_{2}$ from $0{}^{\textrm{o}}$ to $180{}^{\textrm{o}}$ and
obtain ${\rm C}_{{\rm YBE}}$ for each $\theta_{2}$.

\begin{figure}[h]
\centerline{ \includegraphics[scale=0.55]{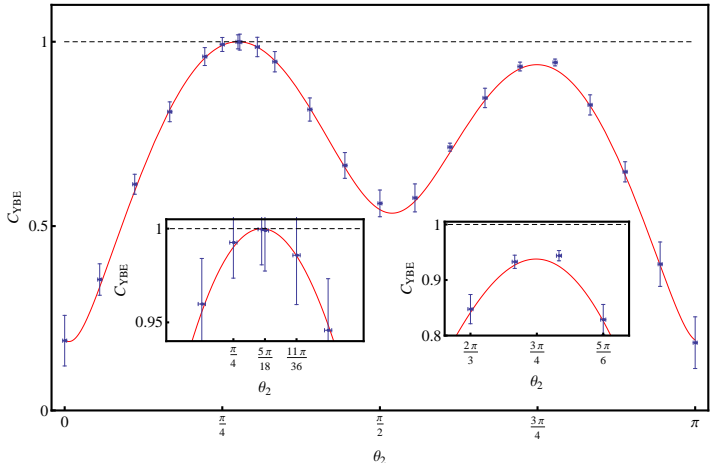}} \caption{(Color online) The curve
of ${\rm C}_{{\rm YBE}}$ versus $\theta_{2}$ where $\theta_{1}$ and
$\theta_{3}$ are kept fixed at $56{}^{\textrm{o}}$ and
$23{}^{\textrm{o}}$, respectively. The dots are the experimental
data while the line is the theoretical curve. The input state is
$|\updownarrow\rangle$, the vertical polarization state. When the
input state is chosen as $|\leftrightarrow\rangle$, the horizontally
polarized state, both experimental and theoretical results are
identical to those for the vertical polarization input state. }
\label{f4}
\end{figure}

The curve of ${\rm C}_{{\rm YBE}}$ versus $\theta_{2}$ for the input
state $|\updownarrow\rangle$ is presented in Fig. \ref{f4}. ${\rm
C}_{{\rm YBE}}$ reaches its maximum value $0.9997\pm0.0237$ when
$\theta_{2}=49.49^{\textrm{o}}$, which is equal 1 within statistical
error. From this one can see that for an arbitrary values of angles,
$\theta_{1}$, $\theta_{2}$ and $\theta_{3}$,
$A(\theta_{3})B(\theta_{2})A(\theta_{1})$ and
$B(\theta_{1})A(\theta_{2})B(\theta_{3})$ are usually not equal.
Only when the three angles satisfy the Lorentz-like relation, Eq.
(\ref{e10}), namely, they satisfy the Yang-Baxter equation, the
operations are equal. This clearly verified the validity of the
Yang-Baxter equation.

For the input state $|\leftrightarrow\rangle$, identical results,
within statistical error, as those for input state
$|\updownarrow\rangle$ are obtained.

\begin{figure}[h]
\centerline{ \includegraphics[scale=0.55]{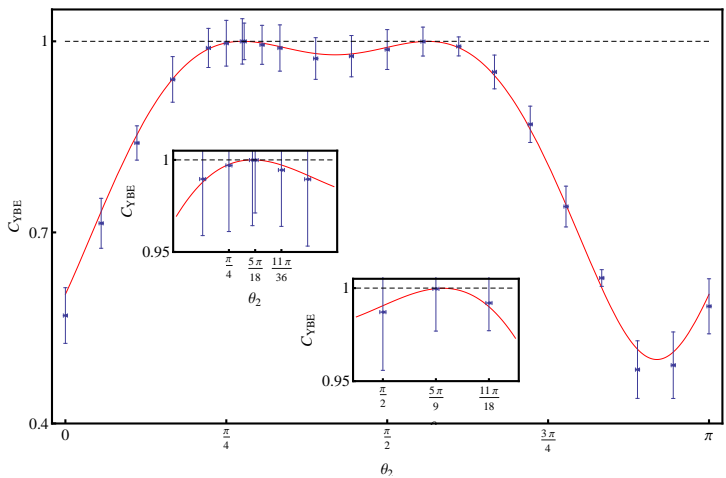}} \caption{(Color online) The curve
of ${\rm C}_{{\rm YBE}}$ versus $\theta_{2}$ where $\theta_{1}$ and
$\theta_{3}$ are kept fixed at $56{}^{\textrm{o}}$ and
$23{}^{\textrm{o}}$, respectively. The dots are the experimental
data while the line is the theoretical curve. The input state is an
elliptically polarized state with the form
$(0.7071-0.5417i)|\updownarrow\rangle-0.4545i|\leftrightarrow\rangle$.
${\rm C}_{{\rm YBE}}$ reaches $0.9999\pm0.0356$ when
$\theta_{2}=49.49^{\textrm{o}}$. }

\label{f5}
\end{figure}

If an input state is an arbitrarily polarized state, say,
$(0.7071-0.5417i)|\updownarrow\rangle$
$-0.4545i|\leftrightarrow\rangle$, the fidelity ${\rm C}_{{\rm
YBE}}$ versus $\theta_{2}$ curve is depicted in Fig. \ref{f5}. One
can see that ${\rm C}_{{\rm YBE}}$ reaches the maximum value $1$
within statistical error when $\theta_{2}=49.49{}^{\textrm{o}}$,
which accords with the theoretical value determined by Eq.
(\ref{e10}); while other $\theta_{2}$ values do not make ${\rm
C}_{{\rm YBE}}$ equal to 1, which implies
$A(\theta_{3})B(\theta_{2})A(\theta_{1})$ and
$B(\theta_{1})A(\theta_{2})B(\theta_{3})$ are not equal when they do
not satisfy Eq. (\ref{e10}). In this case, the Yang-Baxter equation
is not satisfied. Therefore, the two operations are equal only when
the Yang-Baxter equation condition is met. It firmly demonstrates
the validity of the Yang-Baxter equation.

As seen already above, $A(\theta_{3})B(\theta_{2})A(\theta_{1})$ and
$B(\theta_{1})A(\theta_{2})$ $B(\theta_{3})$ are not identical for
arbitrary sets of $\theta_{i}$, $i=1,2,3$. It is clearly manifested
again in the difference between the two curves in Fig. \ref{f4} and
Fig. \ref{f5}, namely the fidelity ${\rm C}_{{\rm YBE}}$ differs for
operations $A(\theta_{3})B(\theta_{2})A(\theta_{1})$ and
$B(\theta_{1})A(\theta_{2})B(\theta_{3})$. They transform the same
input state to different output state for general values of
$\theta_{2}$. The Yang-Baxter equation corresponds to a specific
setting, that is, ${\rm C}_{{\rm YBE}}$ is 1  if three angles
satisfy Eq. (\ref{e10}). In this particular values of $\theta_{1}$
and $\theta_{3}$ displayed in Fig. \ref{f4} and \ref{f5},
$\theta_{2}=49.49^{\textrm{o}}$ satisfies the Yang-Baxter equation,
as they are clearly demonstrated in Fig. \ref{f4} and Fig. \ref{f5}.
The transformations representing the two sides of the Yang-Baxter
equation transform any input state into the same output state,
clearly validating the equality of the Yang-Baxter equation in the
experimental results.

\subsection{Verification for the necessity of Lorentz-like transformation}

We have verified the sufficiency of the Lorentz-like transformation
of parameters $u$ and $v$ in Eq. (\ref{e1}) until now. In term of the
parameters $\theta_{1}$, $\theta_{2}$ and $\theta_{3}$, the transformation
is expressed in Eq. (\ref{e10}). However, It can not be excluded that
other groups of $\theta_{1}$, $\theta_{2}$ and $\theta_{3}$,
which don't satisfy the Lorentz-like transformation, also validate
the Yang-Baxter equation just by the former experimental verification.
It is essential to verify that the Lorentz-like transformation of parameters
$u$ and $v$ in Eq. (\ref{e1}) is the necessary condition to validate the Yang-Baxter equation.
In other words, we need to find out all the relations of $\theta_{1}$,
$\theta_{2}$ and $\theta_{3}$ that make Yang-Baxter equation valid.

Converting it to experiment, we need to record all groups of
$\theta_{1}$, $\theta_{2}$ and $\theta_{3}$ when the fidelity
${\rm C}_{{\rm YBE}}$ reaches 1 within statistical error. To achieve this,
we fixed $\theta_{3}$ first. For each fixed $\theta_{3}$, we kept
$\theta_{2}$ and tune $\theta_{1}$ from 0 to $\pi$ continually to
find out all pairs of $\theta_{1}$ and $\theta_{2}$ that make the
fidelity ${\rm C}_{{\rm YBE}}$  almost 1 (here we choose
the lower bound of ${\rm C}_{{\rm YBE}}$ as 0.9995). The experimental
results show that all groups of $\theta_{1}$, $\theta_{2}$ and $\theta_{3}$
satisfy Eq. (\ref{e10}) when ${\rm C}_{{\rm YBE}}$ attains 1 within statistical error,
i.e. the original spectral parameters $u$ and $v$ satisfy the Lorentz-like transformation.

\begin{figure}[h]
\centerline{\includegraphics[scale=0.8]{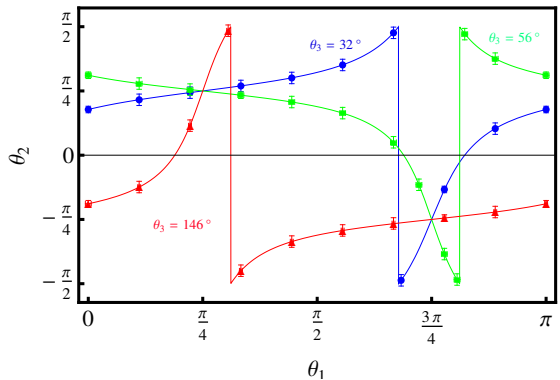}} \caption{(Color online) $\theta_{2}$
versus $\theta_{1}$ curve while fixing $\theta_{3}$. The blue dots,
the green squares and the red triangles are the experimental data
for $\theta_{3}=32{}^{\textrm{o}}$, $56{}^{\textrm{o}}$, and
$146{}^{\textrm{o}}$, while the lines are the corresponding
theoretical results.}
\label{f6}
\end{figure}

To illustrate this, we show three series of our experimental results,
i.e. three curves of $\theta_{2}$ versus $\theta_{1}$ for the corresponding
three fixed $\theta_{3}$ when ${\rm C}_{{\rm YBE}}$ equals to 1 within the
statistical error. We  draw the relation curves of $\theta_{2}$ versus
$\theta_{1}$ for different values of $\theta_{3}$ while keeping
${\rm C}_{{\rm YBE}}>0.9995$ in Fig. \ref{f6}, where
$\theta_{3}$ are fixed at $32^{\textrm{o}}$ (blue dots),
$56^{\textrm{o}}$ (green squares), and $146^{\textrm{o}}$ (red
triangles) respectively. The theoretical figures are also shown in Fig. \ref{f6}
(lines), from which one can see that the experimental results agree
with the theoretical prediction very well. The rich structure of the
Lorentz-like relation is well confirmed in the experiment.

\section{Summary, review and outlook}

The Yang-Baxter equation is directly verified experimentally using
linear quantum optics devices for the first time from the following two aspects.
On one hand, the experiment proved the equality between the two sides of
Yang-Baxter equation if the parameters $\theta_{1}$,
$\theta_{2}$ and $\theta_{3}$ satisfy Eq. (\ref{e10}), which is to $\theta$'s
what Lorentz-like transformation is to the original spectral parameter
$u$ and $v$. It means that the validity of Yang-Baxter equation is guaranteed
sufficiently when the spectral parameters satisfy the Lorentz-like transformation.
On the other hand, we verified that it is also the necessary condition for the
validity of Yang-Baxter equation to make the spectral parameters satisfy the
Lorentz-like transformation. We recorded all groups of the parameters $\theta_{1}$,
$\theta_{2}$ and $\theta_{3}$ that make the fidelity be 1 within the statistical error,
and found out that each group satisfies Eq. (\ref{e10}). In this process, it is fully presented
again that the beautiful structure of the Lorentz-like transformation of the spectral parameters.

Two issues remain open for further
studies for higher dimensional Yang-Baxter equation. One important
issue is the role of entanglement in the Yang-Baxter equation. In
the present experiment, no entanglement is involved. In higher
dimensions, the operations in the Yang-Baxter equation will inevitably bring in
quantum entanglement. It will be an interesting and significant
subject for future study, and consequently the entangling power of
the operations in Yang-Baxter equation, namely the operations in either
sides of the Yang-Baxter equation emerge naturally also an important
topic. The effect of Lorentz-like transformation of the
Yang-Baxter equation is mythical. It drives two independent
operators to become identical as seen from this 2-dimensional
quantum system. In higher dimensions, the operations will be more
complex and entanglement also comes into play. The role of the
transformation relation will be tested and studied in more detail
and in wider aspects.

Discovered from solving problems in many-body systems and statistical models
in the middle of the last century, variety of contexts of Yang-Baxter equation were
revealed and it has been applied to many different area, such as quantum field theory,
statistical mechanics, group theory, and etc. Now, Yang-Baxter equation
is playing an important role in quantum information science which is a thriving area of frontier research.
Using the relation between Bell basis and Yang-Baxter equation enables the investigation of quantum entanglement,
and the relation between anyon and Yang-Baxter equation entails exploring topological
quantum computing. Many interesting applications of the Yang-Baxter equation lies ahead. Yang-Baxter equation not only deserves a direct verification,
like this work, but also merits scientists' continued investigation.

\begin{acknowledgments}
This work was supported by the National Natural Science Foundation
of China (Grant No. 10874 098,11175094), the National Basic Research
Program of China (2009CB929402, 2011CB9216002).
\end{acknowledgments}


\end{document}